%% file: SU4KKHNC.tex
\documentclass[prl,aps,twocolumn,superscriptaddress]{revtex4}
\usepackage{amsmath, amssymb, amsbsy}
\usepackage{color}
\usepackage{array}
\usepackage{txfonts}
\usepackage{bm, bbm}
\usepackage{graphicx}
\usepackage{appendix}
\usepackage{wasysym}
\usepackage{diagbox, multirow}
\usepackage[percent]{overpic}
\usepackage[colorlinks=true,linkcolor=blue,citecolor=blue,urlcolor=blue,bookmarks=false]{hyperref}

\begin{document}
\title{Twisting the Dirac cones of the SU(4) spin-orbital liquid on the honeycomb lattice}

\author{Hui-Ke Jin}
\affiliation {Department of Physics TQM, Technische Universit\"{a}t M\"{u}nchen,  James-Franck-Straße 1, D-85748 Garching, Germany}

\author{W. M. H. Natori}
\affiliation{Institute Laue-Langevin, BP 156, 41 Avenue des Martyrs, 38042 Grenoble Cedex 9, France}

\author{Johannes Knolle}
\affiliation {Department of Physics TQM, Technische Universit\"{a}t M\"{u}nchen,  James-Franck-Straße 1, D-85748 Garching, Germany}
\affiliation{Blackett Laboratory, Imperial College London, London SW7 2AZ, United Kingdom}
\affiliation{Munich Center for Quantum Science and Technology (MCQST), 80799 Munich, Germany}

\begin{abstract}
	By combining the density matrix renormalization group (DMRG) method with Gutzwiller projected wave functions, we study the SU(4) symmetric spin-orbital model on the honeycomb lattice. We find that the ground states can be well described by a Gutzwiller projected $\pi$-flux state with Dirac-type gapless excitations at one quarter filling. Although these Dirac points are gapped by emergent gauge fluxes on finite cylinders, they govern the critical behavior in the thermodynamic limit. By inserting a $\theta=\pi$ spin flux to twist the boundary condition, we can shift the gapless sector to the ground state, which provides compelling evidence for the presence of a gapless Dirac spin-orbital liquid. 
\end{abstract}

\maketitle

{\em\color{blue} Introduction.} 
The search for quantum spin liquids (QSL) is a central problem in condensed matter physics~\cite{Savary2016,Zhou2017,Knolle2019,Broholm2020}. QSLs were originally conceived as phases of the SU(2) Heisenberg model characterized by long-range entanglement instead of local order parameters~\cite{Anderson1973}. However, the ground state of this model often displays long-range order even on geometrically frustrated lattices (e.g., triangular)~\cite{White2007}, which naturally leads to the question of how to induce a quantum disordered liquid phase? One possible theoretical solution is to enhance the symmetry from SU(2) to SU$(N\ge3)$ since the SU$(N)$ antiferromagnetic Heisenberg model in the large-$N$ limit is known to display a QSL ground state~\cite{Affleck1988,Marston1989}. From this exact result follows the principle that quantum fluctuations are amplified by larger symmetries, thus paving a path towards a QSL phase. Besides the theoretical interest, SU($N$) quantum magnetism with $2<N\le10$ has been extensively studied in the context of ultracold atoms in optical lattices~\cite{Hermele2009,Gorshkov2010,Chen2021,ChenReview2021,Chen2022}.

The specific case of the SU$(4)$ Heisenberg model has been the focus of many recent studies due to its relevance in novel solid-state platforms. The system is a special case of a Kugel-Khomskii (KK) model~\cite{Kugel1982,Khaliullin2000,Tokura2020} for Mott insulators retaining two-fold orbital degeneracy, and will be henceforth called the SU(4) KK model. Besides the symmetry principle outlined above, it is often observed in KK models that spin and orbital fluctuations cooperate to increase quantum corrections to order parameters, thus providing another mechanism to exotic quantum phases~\cite{Feiner1997,Reynaud2001}. The SU(4) KK model was initially studied as a simplified model loosely motivated by $e_{g}$ Mott insulators such as LiNiO$_{2}$ ~\cite{Li1998,Pati1998,Penc2003} and Ba$_{3}$Sb$_{2}$CuO$_{9}$~\cite{Corboz2012,Smerald2014}. The first realistic proposal of the SU(4) KK in one-dimensional systems was for Mott insulators with face-sharing octahedra~\cite{Kugel2015}. 
Realistic two-dimensional implementations of the model also have been discussed in $j=3/2$ Mott insulators~\cite{Yamada2018,Natori2018,Yamada2021,Yamada2022} and moir\'{e} materials. In the former case, the honeycomb material ZrCl$_{3}$ is suggested to implement a exchange-frustrated model which can be mapped into an SU(4) symmetric Hubbard model by SU(4) gauge transformations~\cite{Yamada2018} or sublattice-dependent pseudospin rotations~\cite{Natori2018}. The SU(4) KK systems are also proposed to be realized in the correlated insulating phase of moir\'{e} materials, specially for those systems whose low energy degrees of freedom form triangular lattices~\cite{Xu2018,Po2018,Zhang2019,Wu2019,Schrade2019,Zhang2021}. For instance, the SU(4) KK model on the honeycomb lattice was initially put forward as a good starting point for magic-angle twisted bilayer devices~\cite{Yuan2018,Venderbos2018,Natori2019}, which was later ruled out because of the impossibility of defining Wannier orbitals out of graphene Dirac points~\cite{Po2018}. However, more recent ab initio studies suggest that the honeycomb SU(4) KK can be relevant in moir\'{e} systems on transition metal dichalcogenide bilayers~\cite{Angeli2021,Xian2021}.

For the past decades, a great effort has been devoted to investigating quantum phases in SU(4) symmetric quantum magnets. In one dimension, the SU(4) KK model is integrable and has gapless excitations~\cite{Sutherland1975,Li1999}, which is described by the SU(4)$_1$ Wess-Zumino-Witten (WZW) conformal field theory (CFT)~\cite{Affleck1986,Azaria1999,fuhringer2008}. The ground state of the SU(4) KK model on a two-leg ladder is an SU(4)-singlet plaquette valence-bond crystal breaking the translational invariance~\cite{Bossche2001,Chen2005,Weichselbaum2018}. 
In two dimensions, this model is less well studied and many important questions remain open, as can be illustrated for the case of the honeycomb lattice. Earlier infinite Projected Entangled Pair States (iPEPS) and variational Monte Carlo (VMC) studies indicated that the ground state of SU(4) KK model is a Dirac-type spin-orbital liquid~\cite{Corboz2012}. An extended version of the Lieb-Schultz-Mattis theorem has been extensively studied for this model~\cite{Yamada2018,Yamada2021,Yamada2022} and allows for such a gapless QSL. 
However, a recent investigation focusing on the specific heat indicates a gapped QSL with topological order~\cite{Yamada2022}. Overall, the ground state of the SU(4) KK model on the honeycomb lattice still remains elusive.

In this work, we revisit the SU(4) KK model on the honeycomb lattice. We utilize the  density matrix renormalization group (DMRG) method~\cite{White1992,White1993} (up to bond dimension $\chi=14000$) to investigate the ground states on finite cylinders. A newly developed methodology~\cite{Tu2020,Jin2020,petrica2020,aghaei2020,Jin2020_2,Jin2022} allows us to exploit the Gutzwiller projected wave functions to characterize the ground states obtained by DMRG. With extensive numerical efforts and analytical analyses, we find that the ground states can be well described by a Gutzwiller projected $\pi$-flux state with Dirac-type gapless excitations by verifying the wave function fidelity. Although these Dirac points are gapped by emergent gauge fluxes on finite cylinders, we argue that they can manifest themselves in the 2D limit where the effect of gauge fluxes are negligible. Remarkably, we can twist the Dirac points to the ground state sector by inserting a $\theta=\pi$ spin flux. Usually, 2D DMRG algorithm can only work well on cylindrical geometries, which might provide misleading information, known as one of the biggest disadvantages. We emphasize that the combination of DMRG and Gutzwiller projected state can make up for this shortcoming and provides a promising way for approaching the 2D limit.


\begin{figure}[t!]
    \includegraphics[width=\linewidth]{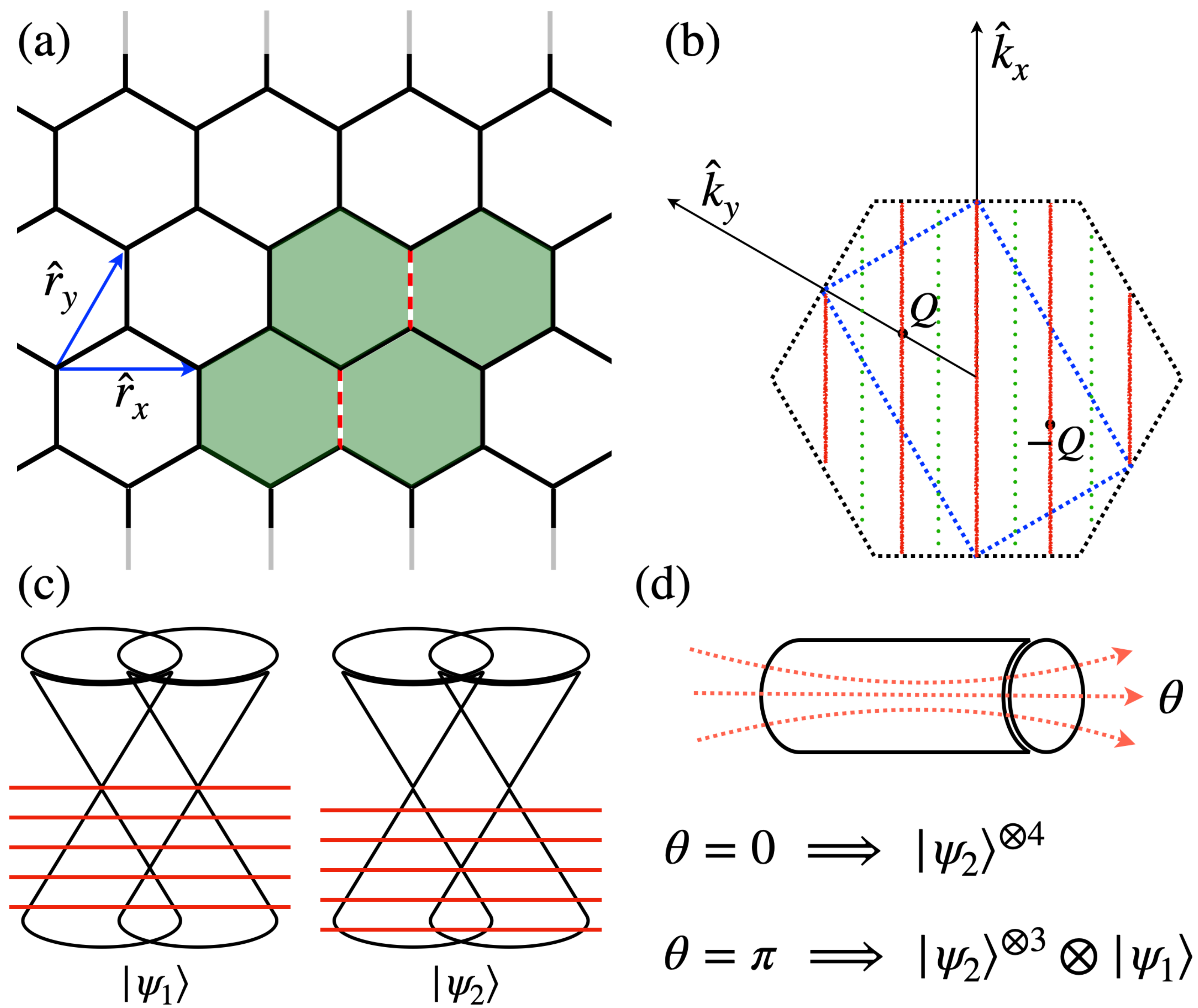}
    \caption{ (a) YC-8 cylinder for the honeycomb lattice with unit vector $\hat{r}_x$ and $\hat{r}_y$, where the half transparent bonds indicate twisted boundary conditions for SU(4) spins. The colored area denotes a unit cell of the $\pi$-flux state defined in Eq.~\eqref{eq:Hpi}, where the the solid black bonds and red dashed bonds are $+1$ and $-1$ hopping terms, respectively. (b) The first Brillouin zone (dashed hexagon) of the honeycomb lattice and the folded Brillouin zone (blue dashed rectangle) for the $\pi$-flux state. The two Dirac points of the $\pi$-flux state are at $\pm{}Q=\pm{}(\pi/2,\pi/2)$. Red rough lines (green dashed lines) represent momenta on YC-8 cylinders that are allowed by PBC (APBC) along the $y$ direction for partons. (c) The gapless sector $|\psi_1\rangle$ and gapped sector $|\psi_2\rangle$ of the $\pi$-flux state. It is easy to see that with the same number of allowed momenta, the gapped sector has lower energy than the gapless sector by $\sim{}v_pL_xL_y^2$, where $v_p$ is the Fermi velocity of partons. (d) The honeycomb SU(4) KK model on a finite cylinder without spin flux ($\theta=0$) is gapped due to emergent gauge flux, while it has exact gapless Dirac points by inserting a spin flux $\theta=\pi$. } \label{fig:latt}
\end{figure}

{\em\color{blue} Model.}
We focus on the SU(4) KK model on the honeycomb lattice [see Fig.~\ref{fig:latt}(a)] defined by
\begin{equation}
	H = \frac{1}{2}\sum_{\langle i j \rangle} \left(  {\bm \sigma}_i \cdot {\bm
		\sigma}_j +1 \right) \left(  {\bm \tau}_i \cdot {\bm \tau}_j +
	1 \right),    \label{eq:SU4Model}
\end{equation}
where $\langle{}ij\rangle$ denotes the nearest-neighbor (NN) bonds and $\bm{\sigma}$ ($\bm{\tau}$) represents Pauli matrices for spin (orbital) degrees of freedom. The spin-orbital system has a four-dimensional local basis denoted by $|m\rangle$ with $m=1,2,3,4$. Indeed, the local Hilbert space can form the fundamental representation of the SU(4) Lie algebra, and the Hamiltonian \eqref{eq:SU4Model} can be rewritten in an SU(4)-invariant form as
\begin{equation*}
H=\sum_{\langle{}ij\rangle}\left[\frac{1}{2}\sum_{\alpha=1}^{3}\lambda^\alpha_i\lambda^\alpha_j+\sum_{m=1}^{4}\sum_{n>m}\left(\lambda^{mn}_{i}\lambda^{nm}_j+{\rm H.c.}\right)\right],
\end{equation*}
where the above $4\times{}4$ matrices $\lambda$'s are the fifteen SU(4) generators. Explicitly, $\lambda^\alpha_i$ ($\alpha=1,2,3$) are three Cartan generators and $\lambda^{mn}=|m\rangle\langle{}n|$ ($n\neq{}m$ and $m,n=1,2,...,4$) are twelve raising ($m>n$) and lowering ($m<n$) operators of the SU(4) Lie algebra~\cite{appendix}.

{\em\color{blue} Fermionic parton construction.}
For SU($N$) quantum magnets, 
parton constructions are usually considered an efficient method to derive effective theories and to construct variational wave functions. Here, we adopt this strategy by introducing the SU(4) fermionic parton representation~\cite{Wang2009,Corboz2012,Xu2020,Yamada2021,Zhang2021,Jin2022_2},
\begin{align}
	\lambda^{\alpha}_{i} \rightarrow \bm{f}^\dagger_i\lambda^{\alpha}\bm{f}_i,
	~\lambda^{mn}_{i}\rightarrow\bm{f}^\dagger_i\lambda^{mn}\bm{f}_i,
\end{align}
where $\bm{f}^\dag_{i}=(f^\dag_{i,1},f^\dag_{i,2},f^\dag_{i,3},f^\dag_{i,4})$ is a four-component vector of the creation operators for fermionic partons ($\bm{f}_{i}$ denote annihilation operators). This parton representation will enlarge the Hilbert space and introduces a redundant $U(1)$ gauge structure~\cite{Wen02}. In order to obtain the physical wave functions, one has to enforce the constraint $\sum_{m=1}^4 f^\dagger_{i,m}f_{i,m}=1$ by projecting the local Hilbert space onto single occupancy.

One can exploit the parton representation to perform a mean-field decomposition of the original Hamiltonian to obtain an effective Hamiltonian of partons. The effective Hamiltonian is usually quadratic in partons, which fully determines the trail mean-field ground states and corresponding low-energy excitations. For instance, the uniform $\pi$-flux state proposed in Ref.~\cite{Corboz2012} is the ground state of the following effective Hamiltonian:
\begin{equation}
	H_{\rm \varhexagon =\pi{}} = -\sum_{m=1}^4\sum_{\langle ij \rangle} t_{ij}\left( f^\dag_{i,m}f_{j,m} + f^\dag_{j,m}f_{i,m}\right). \label{eq:Hpi}
\end{equation}
Here, $t_{ij}=\pm{}1$ and their signs are indicated in Fig.~\ref{fig:latt}(a). Although the parton Hamiltonian breaks the translational symmetry along the $x$-direction by doubling the unit cell, this broken symmetry is restored after Gutzwiller projection~\cite{Wen02}. 
At one quarter filling, the band structure of Eq.~\eqref{eq:Hpi} for each flavor has two gapless Dirac cones at momenta $\pm{Q}=\pm(\pi/2,\pi/2)$ in the folded Brillouin zone, as shown in Fig.~\ref{fig:latt}(b). Therefore, after counting all four flavors, overall there are eight Dirac cones in the parton mean-field level.

Next, we study the ground state of the original Hamiltonian~\eqref{eq:SU4Model} and various effective parton Hamiltonians with matrix product state (MPS) techniques. For the effective parton Hamiltonians, we adopt the newly developed method~\cite{Tu2020,Jin2020,Jin2022}, rather than conventional DMRG, to directly convert the mean-field ground state into the MPS form. Then we can easily implement the Gutzwiller projection upon the parton MPS to obtain the physical many-body wave function.
In their MPS forms, these many-body wave functions serve as variational ansatz for Hamiltonian~\eqref{eq:SU4Model}, and meanwhile can be utilized to initialize DMRG calculations for Hamiltonian~\eqref{eq:SU4Model}, which greatly improves the convergence of the DMRG algorithm~\cite{Jin2020_2}.

To perform MPS-related calculations on 2D lattices, we place the systems on cylindrical geometries. We work with the YC-$2L_y$ cylinders with circumference $L_y$, where the periodic boundary condition (PBC) for spin-orbital degrees of freedom is imposed along the $y$ direction [corresponding to $\hat{r}_y$ in Fig.~\ref{fig:latt}(a)] while the $x$ direction with length $L_x$ is left open. The mapping of the YC cylinders can be found in the Supplemental Material~\cite{appendix}.
`
Note that the fermionic partons are coupled to the emergent $U(1)$ gauge field which can lead to a global gauge flux, $\Phi$, through the cylinder. A time-reversal invariant state requires either a $\Phi=0$ or $\Phi=\pi$, corresponding to PBC or anti-periodic boundary condition (APBC) for parton degrees of freedom, respectively. For finite cylinders, those PBC and APBC correspond to different ways of cutting the Brillouin zone, see Fig.~\ref{fig:latt}(b). Crucially, for a cylinder with $L_y$ mod 2 even (odd), the allowed momenta can cut the Dirac points of the $\pi$-flux state exactly when $\Phi=0$ ($\Phi=\pi$). Because the partons with different flavors are decoupled from each other in our $U(1)$ parton theories, we can treat $\Phi_m$ separately as discrete parameters to tune the appearance and disappearance of the Dirac points for each flavor $m=1,2,3,4$. 

\begin{figure}[t!]
    \centering
    \includegraphics[width=\linewidth]{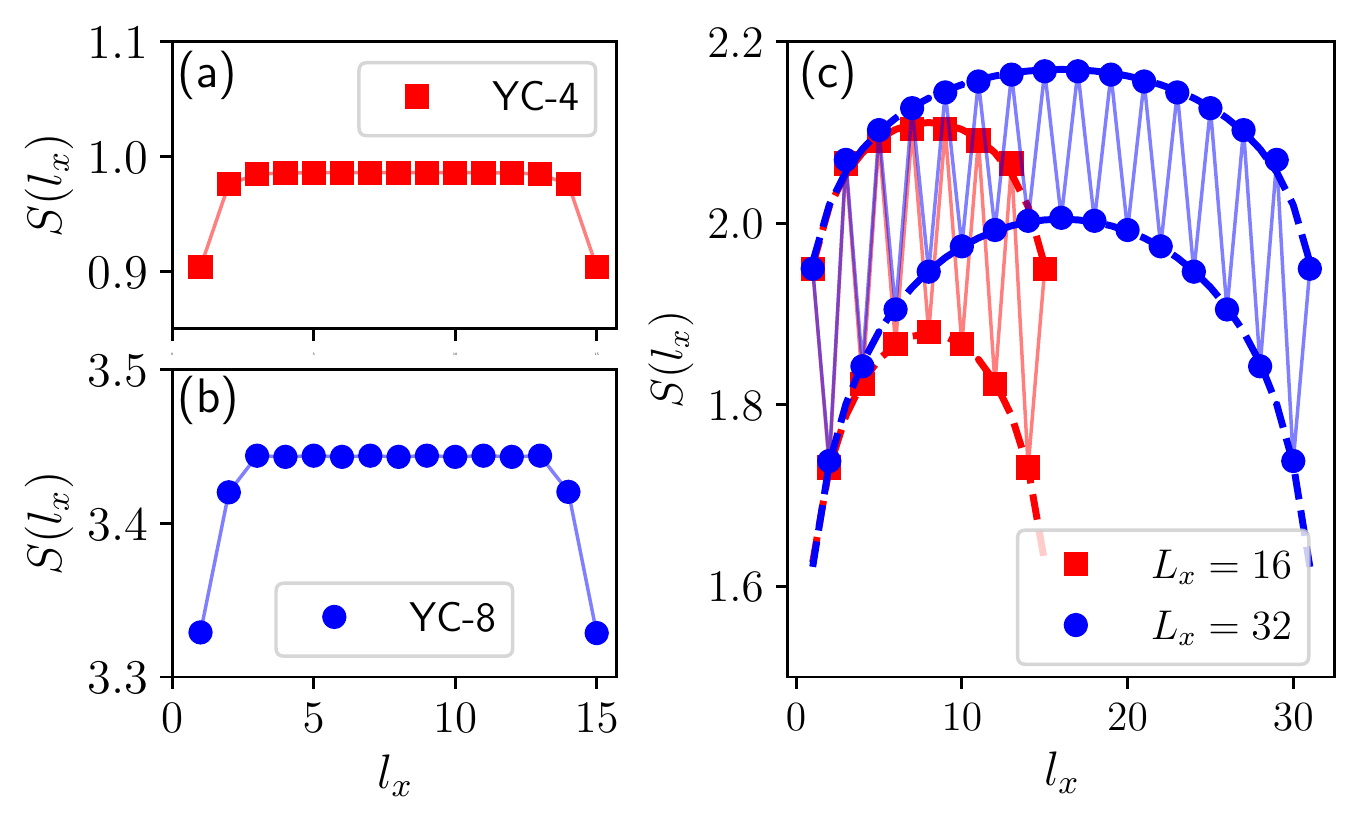}
    \caption{The bipartite EE $S(l_x)$ versus subsystem length $l_x$ for the ground states of Hamiltonian~\eqref{eq:SU4Model} (a) on YC-$4$ cylinders, (b) YC-$8$ cylinders, and (c) on twisted YC-$4$ cylinders. The quickly saturated $S(l_x)$ in (a) and (b) indicate gapped states. The extracted central charge for the lower branches of $S(l_x)$ in (c) is $c\approx{}0.93$ for $L_x=16$ and $c\approx{}0.99$ for $L_x=32$.}
    \label{fig:Slx}
\end{figure}

{\em\color{blue} Results.}
Thanks to the combinative methods, i.e., DMRG and Gutzwiller projected wave function, we have obtained the ground state of the honeycomb KK model~\eqref{eq:SU4Model} (denoted by $|\Psi_{\rm G}\rangle$ hereafter) on YC-$4$ and YC-$8$ cylinders. In order to characterize it, we first calculate the SU(4) spin correlation function, $\Lambda_{ij}\equiv\langle\bm{\mathcal{S}}_i\cdot\bm{\mathcal{S}}_j\rangle$, where $\bm{\mathcal{S}}_i=(\sigma^a_i,\tau^b_i,\sigma^a_i\tau^b_i)_{a,b=x,y,z}$. The values of $\Lambda_{ij}$ on the NN bonds, i.e., the local energy densities, exhibit a uniform pattern in the bulk~\cite{appendix}, indicating that there is no translational symmetry breaking in both directions. This important observation rules out the possibility of stripy states~\cite{Jin2022_2} and valence-bond crystals~\cite{Chen2021}. 

To address the ``gapped or gapless'' issue of Hamiltonian~\eqref{eq:SU4Model}, we study the von Neumann entanglement entropy (EE) of $|\Psi_{\rm D}\rangle$. By treating a cylinder as a quasi-1D chain with a column unit cell of $2L_y$ spins, we can divide the whole cylinder into two parts, i.e., the first $l_x$ column unit cells and the remaining $L_x-l_x$ ones. Then we are able to calculate the bipartite EE $S(l_x)$ as a function of $l_x$. For a quasi-1D gapless state, $S(l_x)$ is described by a CFT, satisfying the following scaling law~\cite{Holzhey1994,Vidal2003,Cardy2009}:
\begin{equation}
	S(l_x)=\frac{c}{6}\log\left(\frac{L_x}{\pi}\sin\frac{\pi{}l_x}{L_x}\right) +\gamma \label{eq:CFTEE}
\end{equation}
with central charge $c>0$. As shown in Figs.~\ref{fig:Slx}(a) and (b), $S(l_x)$ saturates quickly with $l_x$ and changes little with bond dimension $\chi$ for both YC-$4$ and YC-$8$ cylinders, respectively, indicating that the ground state of Hamiltonian~\eqref{eq:SU4Model} is gapped on finite cylinders. We emphasize that this gapped signature is somehow consistent with the results of Ref.~\cite{Yamada2022}.

However, for finite cylinders, a gapped ground state is not necessarily in conflict with the $\pi$-flux state, since the allowed momenta along $y$ directions might not exactly cut the Dirac points of Eq.~\eqref{eq:Hpi} due to the emergent gauge flux $\Phi_m$ [for instance, see Fig.~\ref{fig:latt}(b)]. 
Indeed, this emergent gauge flux $\Phi_m$ can be extracted by verifying the wave function fidelity $F=|\langle\Psi_{\rm G}|\Psi_{\rm D}\rangle|$ with $|\Psi_{\rm G}\rangle$ the Gutzwiller projected $\pi$-flux state. 
As listed in Table~\ref{tab:fidelity}, for a YC-$4$ cylinder with length $L_x=8$, we find that $F\approx{}0.985$ for a projected $\pi$-flux state with $\Phi_m=0$ (zero Dirac point cut) and $F\approx{}0.014$ for that with $\Phi_m=\pi$ (two Dirac points for each flavor). Similar results can be obtained on YC-$8$ cylinder, in which $F\approx{}0.907$ for a $\pi$-flux state without Dirac point ($\Phi_m=\pi$) and $F\approx{}0.216$ with eight Dirac points ($\Phi_m=0$). These remarkably large wave function fidelities indicate that the ground state of Hamiltonian~\eqref{eq:SU4Model} indeed is a $\pi$-flux state, but avoids cutting the Dirac points on finite cylinders. Notice that the number of Dirac points is naively counted at the mean-field level, which usually is reduced after Gutzwiller projection. We also find that the quality of  zero-flux states with uniform hoppings is poor as it has a negligible wave function fidelity.

\begin{table}[t!]
    \centering
    \caption{The wave function fidelity between $|\Psi_{\rm D}(\theta)\rangle$ obtained by DMRG and Gutzwiller projected $\pi$-flux state $|\Psi_{\rm G}\rangle$. Here $\theta=0$ and $\theta=\pi$ corresponds to $|\Psi_{\rm{}D}\rangle$ on usual PBC and twisted boundary conditions, respectively. The number of Dirac cones for $|\Psi_{\rm G}\rangle$ is controlled by gauge flux $\Phi_m$.}
    \renewcommand\arraystretch{1.5}
    \setlength{\arrayrulewidth}{0.3mm}
    \setlength{\tabcolsep}{2.0ex}
    \begin{tabular}{cccc}
    \hline
    \hline
     \multicolumn{4}{c}{$|\langle\Psi_{\rm G}|\Psi_{\rm D}\rangle|$ on YC-$4$ cylinder with $L_x=8$ }\\
    \hline
    &  8 Dirac points & 0 Dirac point           & 2 Dirac points\\
    \hline
    $\theta=0$ & 0.014 & 0.985  & 0.192 \\
    $\theta=\pi$ & 0.022  & 0.213 & 0.904\\
    \hline
    \hline
     \multicolumn{4}{c}{$|\langle\Psi_{\rm G}|\Psi_{\rm D}\rangle|$ on YC-$8$ cylinder with $L_x=4$ }\\
    \hline
    &  8 Dirac points & 0 Dirac point           & 2 Dirac points\\
    \hline
    $\theta=0$ & 0.286 & 0.907  & 0.320 \\
    $\theta=\pi$ & 0.229  & 0.344 & 0.893\\
    \hline
    \hline
    \end{tabular}
    \label{tab:fidelity}
\end{table}

As illustrated in Fig.~\ref{fig:latt}(c), by filling the single-particle states below the Dirac point, the gapped sector can gain energy of $\delta{}E_{L_y}\sim{}v_pL_x/L_y^2$ with $v_p$ the Fermi velocity of partons. Therefore, the ground state $|\Psi_{\rm D}\rangle$ always energetically favors the emergent gauge flux $\Phi$ which avoids cutting the Dirac points, but in the thermodynamic limit, where this energy difference $\delta{}E_{L_y}\rightarrow0$ when $L_y\rightarrow\infty$, the gapped and gapless sectors are degenerate. Our results thus strongly support the gapless spin-orbital liquid senario obtained by previous iPEPS and VMC studies~\cite{Corboz2012}.

{\em\color{blue} Twist boundary condition.}
The gapless nature of Hamiltonian~\eqref{eq:SU4Model} can still be revealed even on finite cylinders. Following the strategy introduced in Refs.~\cite{Haldane95,Gong2014,He2017}, we consider generalized PBC in which the SU(4) spin operators acquire a twisted boundary condition, namely, taking the raising operators associated with the $|m=1\rangle$ local states as $\lambda^{n1}_{i+L_y\hat{r}_y}={}e^{i{}\theta/2}\lambda^{n1}_{i}$ $(n=2,3,4)$ [corresponding lowering operators: $\lambda^{1n}=(\lambda^{n1})^\dag$]~\cite{appendix}. Here $\theta$ is a so-called spin flux in the cylinder which reduces the SU(4) symmetry into a U(1)$^{\otimes{}3}$ one. This effect is of order $1/L_y$ and we expect that it will not have a significant effect in the bulk. 
To preserve the time-reversal symmetry, we only consider a $\theta=\pi$ spin flux besides the trivial one $\theta=0$. By choosing a proper gauge, a $\theta=\pi$ spin flux modifies the exchange terms on the NN bonds $\langle{}ij\rangle$ {\em only} along the $y$ (periodic) boundary [see Fig.~\ref{fig:latt}(a)] as
\begin{equation}
\sum_{n=2}^{4}\left(\lambda^{n1}_{i}\lambda^{1n}_{j}+{\rm{}H.c.}\right)\longrightarrow{}-\sum_{n=2}^{4}\left(\lambda^{n1}_{i}\lambda^{1n}_{j}+{\rm{}H.c.}{}\right),    \label{eq:twist}
\end{equation}
and leaves the other terms in Eq.~\eqref{eq:SU4Model} {\em unchanged}. Since the ground state of the original Hamiltonian~\eqref{eq:SU4Model} is a $\pi$-flux state, we expect that this $\theta=\pi$ spin flux can pump the $m=1$ flavor partons to exactly cut the Dirac point.

We denote the ground state of the Hamiltonian~\eqref{eq:SU4Model} with twisted boundary obtain by DMRG by $|\Psi_{\rm D}(\theta=\pi)\rangle$. The entanglement entropy $S(l_x)$ of $|\Psi_{\rm D}(\theta=\pi)\rangle$ splits into two branches, where the upper (lower) branch corresponds to an odd (even) $l_x$, see Fig.~\ref{fig:Slx}(c). The even-odd oscillations are induced by the open boundary condition along the $x$ direction~\cite{Laflorencie2006}. Both branches of $S(l_x)$ on the YC-$4$ cylinders with $\theta=\pi$ are well described by Eq.~\eqref{eq:CFTEE},  and the extracted central charge, $c\approx{}1$, for the lower branch is consistent with the results predicted by the CFT of a single-component massless fermion system.

By treating the gauge flux coupling to the $m=1$ partons and the other gauge fluxes separately, we can prepare the projected parton states which efficiently characterize $|\Psi_{\rm D}(\theta=\pi)\rangle$. For instance, by choose $\Phi_1=\pi$ and $\Phi_m=0$ ($\Phi_1=0$ and $\Phi_m=\pi$) with $m=2,3,4$, again we can prepare a parton state on a YC-$4$ (YC-$8$) cylinder which only contains two Dirac points [see Fig.~\ref{fig:latt}(d)]. 
We find that the wave function fidelities between $|\Psi_{\rm D}(\theta=\pi)\rangle$ and those projected states containing two Dirac points are $F\approx{}0.907$ on YC-$4$ cylinder and $F\approx{}0.893$ on YC-$8$ cylinder, as listed in Table~\ref{tab:fidelity}. These remarkably high fidelities further provide strong evidence that the gapless state $|\Phi_m(\theta=\pi)\rangle$ is still a $\pi$-flux state.

{\em\color{blue} Discussion.}
In summary, we have studied the SU(4) KK model on the honeycomb lattice by a novel DMRG method built on Gutzwiller projected parton wave functions. We provide strong evidence for a gapless quantum spin-orbital liquid ground state with Dirac-type excitations in the 2D limit, which is efficiently described by a $\pi$-flux state. The Dirac points of this spin-orbital liquid are gapped due to an emergent gauge flux on finite cylinders. Using a parton ansatz, we have shown how it can be revealed by inserting a $\theta=\pi$ spin flux to twist the boundary condition. We expect that this $\pi$-flux Dirac liquid state is stable beyond the quasi-1D geometry and serves as an excellent ground-state candidate for the honeycomb SU(4) KK model in the thermodynamic limit, in agreement with Ref.~\cite{Corboz2012}. 

The MPS representation of Gutzwiller projected wave functions provides a powerful tool for directly computing several quantities of interest such as wave function fidelities. Our results demonstrate the advantages of the method, i.e., to circumvent the strong finite-size effects of standard 2D DMRG simulations. Besides being a useful tool for probing topological quantities~\cite{Haldane95,Tu2013,Gong2014,He2017}, twisting boundary conditions can also be used to uncover possible parton Fermi surface states gapped by emergent gauge flux, e.g., the proposed deformed parton Fermi surface in the SU(4) KK model on the triangular lattice~\cite{Jin2022_2}.

For future works, it would be interesting to map out the phase diagram of the KK model near the SU(4) symmetric point but including experimentally relevant perturbing interactions. For instance, the Dirac spin-orbital state might server as a critical point for several nearby phases such as a chiral spin liquid induced by three-body interactions~\cite{Lauchli2016,Zhang2021}. Another intriguing prospect is to observe/determine signatures of this Dirac spin-orbital liquid in real materials such as ZrCl$_{3}$~\cite{Yamada2018,Natori2018}. As a promising trial wave function, the $\pi$-flux state is an excellent starting point for predicting various dynamical spectral functions which is currently beyond the capabilities of MPS-based methods.


\begin{acknowledgements}
	{\em\color{blue} Acknowledgments.} We thank Frank Pollmann for helpful discussions. H.-K.J. especially thanks Hong-Hao Tu for his comments and suggestions. W.M.H.N. acknowledges R. Pereira, E. Andrade, and R. Nutakki for participation in related projects~\cite{Natori2018,Natori2019}. H.-K.J. is funded by the European Research Council (ERC) under the European Unions Horizon 2020 research and innovation program (grant agreement No. 771537). J.K. acknowledges support from the Imperial-TUM flagship partnership. The research is part of the Munich Quantum Valley, which is supported by the Bavarian state government with funds from the Hightech Agenda Bayern Plus. The numerical simulations in this work are based on the GraceQ project~\cite{graceq}.
\end{acknowledgements}

\bibliography{SU4KKHNC}
\clearpage
\input{KKhoneycombSM}

\end{document}

%% file: KKhoneycombSM.tex


\begin{widetext}

\begin{center}
{\Large\bf Supplemental material for ``Twisting the Dirac cones of the SU(4) spin-orbital liquid on the honeycomb lattice''}
\end{center}




\setcounter{equation}{0}
\setcounter{figure}{0}
\setcounter{table}{0}
\setcounter{section}{0}

\renewcommand{\theequation}{S\arabic{equation}}
\renewcommand{\thefigure}{S\arabic{figure}}
\renewcommand{\thetable}{S\Roman{table}}
\renewcommand{\thesection}{S\Roman{section}}

This Supplemental Material provides technique details about (i) the one-dimensional path for YC cylinders, (ii) DMRG results, and (iii) basic SU(4) algenra and spin flux.

\section{One-dimensional path}
In order to carry out the MPS-related methods, we first define the ordering of lattice sites, which can be implemented  by assigning an integer  $\tilde{j}=1,\cdots,N$ to each lattice site, where $N$ is the number of lattice sites. Here we adopt a site-labeling scheme for honeycomb lattice on the YC-$2L_y$ cylinders, as illustrated in Fig.~\ref{fig:YC} 

\begin{figure}[tbhp]
\includegraphics[width=0.65\linewidth]{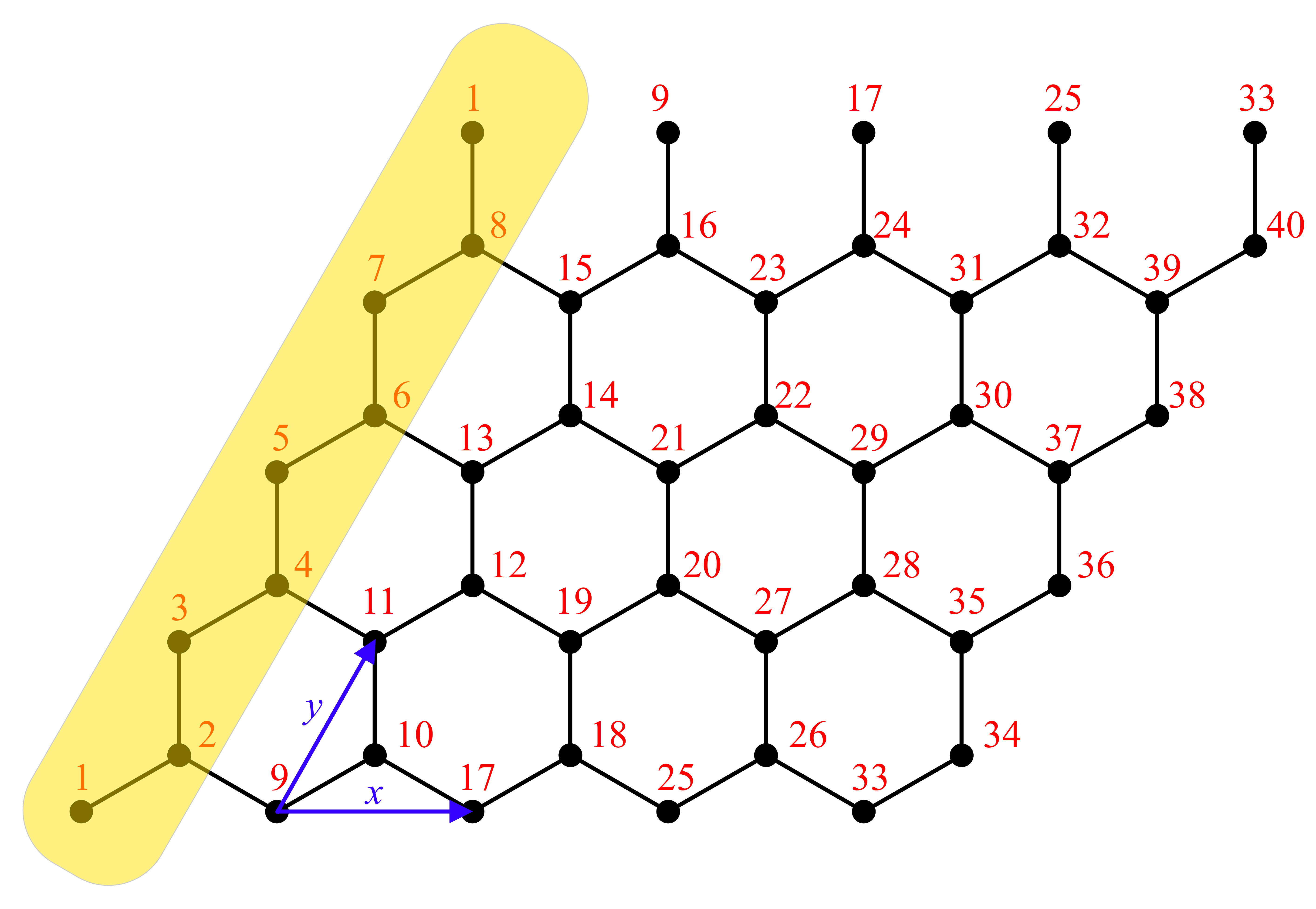}
\caption{Schematics of the labeling scheme for a honeycomb lattice on a YC-$8$ cylinder with length $L_x=5$. The number of total lattice sites are $2L_yL_x=2\times{}8\times{}5=40$. Notice that the periodic boundary condition has been imposed along the $y$ direction. The yellow shadow indicates one column unit cell for Eq.~(4) in the main text.}
\label{fig:YC}
\end{figure}

\section{DMRG results}

\subsection{Ground-state energy}
\begin{figure}[!h]
	\includegraphics[width=0.4\linewidth]{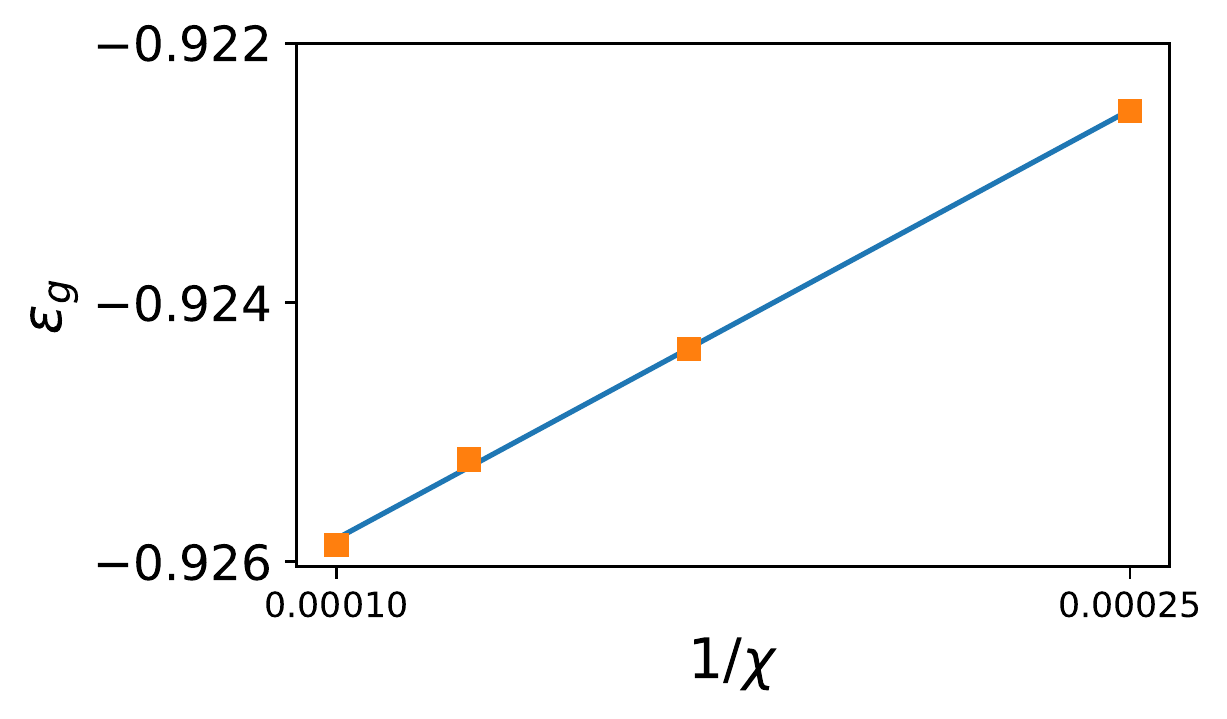}
	\caption{Per site ground-state energy $\epsilon_g$ as a function of DMRG bond dimension $\chi$ on a YC-8 cylinder with length $L_x=16$. }
	\label{fig:engs_chi}
\end{figure}
To verify the convergence of the ground states obtained by DMRG, we study the scaling behavior of the per site ground-state energy, $\epsilon_g$, as a function of DMRG bond dimension $\chi$ on a YC-$8$ cylinder with length $L_x=16$. As shown in Fig.~\ref{fig:engs_chi}, $\epsilon_g$ scales linearly with inverse bond dimension $1/\chi$, indicating a faithful convergence of DMRG. Here we initialized our DMRG simulations with a random MPS. Similar energy can be obtained by the DMRG initialized with Gutzwiller $\pi$-flux states at the same bond dimension, where $\epsilon_g\approx{}-0.926$ at $\chi=10000$ on a YC-$8$ cylinder with length $L_x=16$.

\subsection{$\Lambda_{ij}$ on NN bonds}
The SU(4) spin-spin correlation function $\Lambda_{ij}$ has been defined in the main text as $$\Lambda_{ij}\equiv\langle\bm{\mathcal{S}}_i\cdot\bm{\mathcal{S}}_j\rangle,$$ where $\bm{\mathcal{S}}_i=(\sigma^a_i,\tau^b_i,\sigma^a_i\tau^b_i)_{a,b=x,y,z}$. On the NN bonds, $\Lambda_{ij}$ are just the bond energy for Hamiltonian in Eq.~(1) in the main text.

\begin{figure}[!h]
	\centering
	\includegraphics[width=0.9\linewidth]{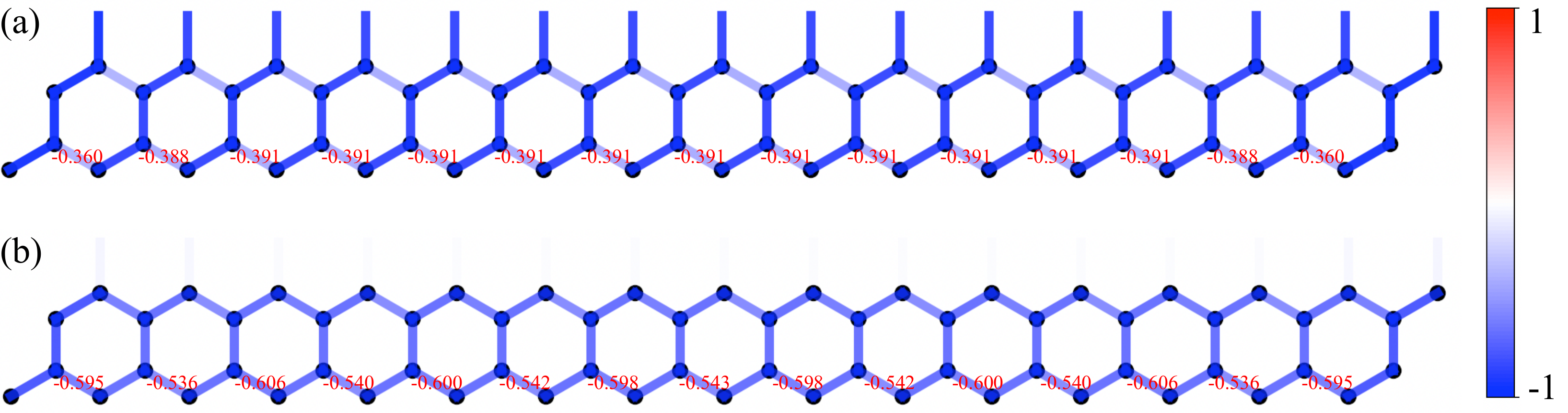}
	\caption{The expectation values of correlation function $\Lambda_{ij}$  on the NN bonds (a) on a YC-$4$ cylinder with spin flux $\theta=0$ and (b) on a twisted YC-$4$ cylinder with $\theta=\pi$. The colors of the bonds represent the expectation values.}\label{fig:energy_bond_yc4}
\end{figure}

In Fig.~\ref{fig:energy_bond_yc4}, we show $\Lambda_{ij}$ on the NN bonds for both twisted and untwisted YC-$4$ cylinders.  In the bulk of Fig.~\ref{fig:energy_bond_yc4}(a), $\Lambda_{ij}$ are very uniform, indicating that the ground state preserves translational invariant symmetry on a YC-$4$ cylinder with spin flux $\theta=0$.  On the YC-$4$ cylinder with nonzero spin flux $\theta=\pi$, it seems that the ground state exhibits ``translational symmetry breaking'' in viewing from $\Lambda_{ij}$, as shown in Fig.~\ref{fig:energy_bond_yc4}(b). However, this two-fold oscillation of $\Lambda_{ij}$ is smeared out in the deep bulk of the cylinder, which is an open-boundary effect. We also emphasize that this two-fold oscillation can also be observed in the uniform Gutzwiller projected $\pi$-flux states on cylinders, which, however, definitely is a translational invariant state.  Here the MPS bond dimension $\chi=4000$ for computing  Fig.~\ref{fig:energy_bond_yc4}.

\begin{figure}[!h]
	\centering
	\includegraphics[width=0.9\linewidth]{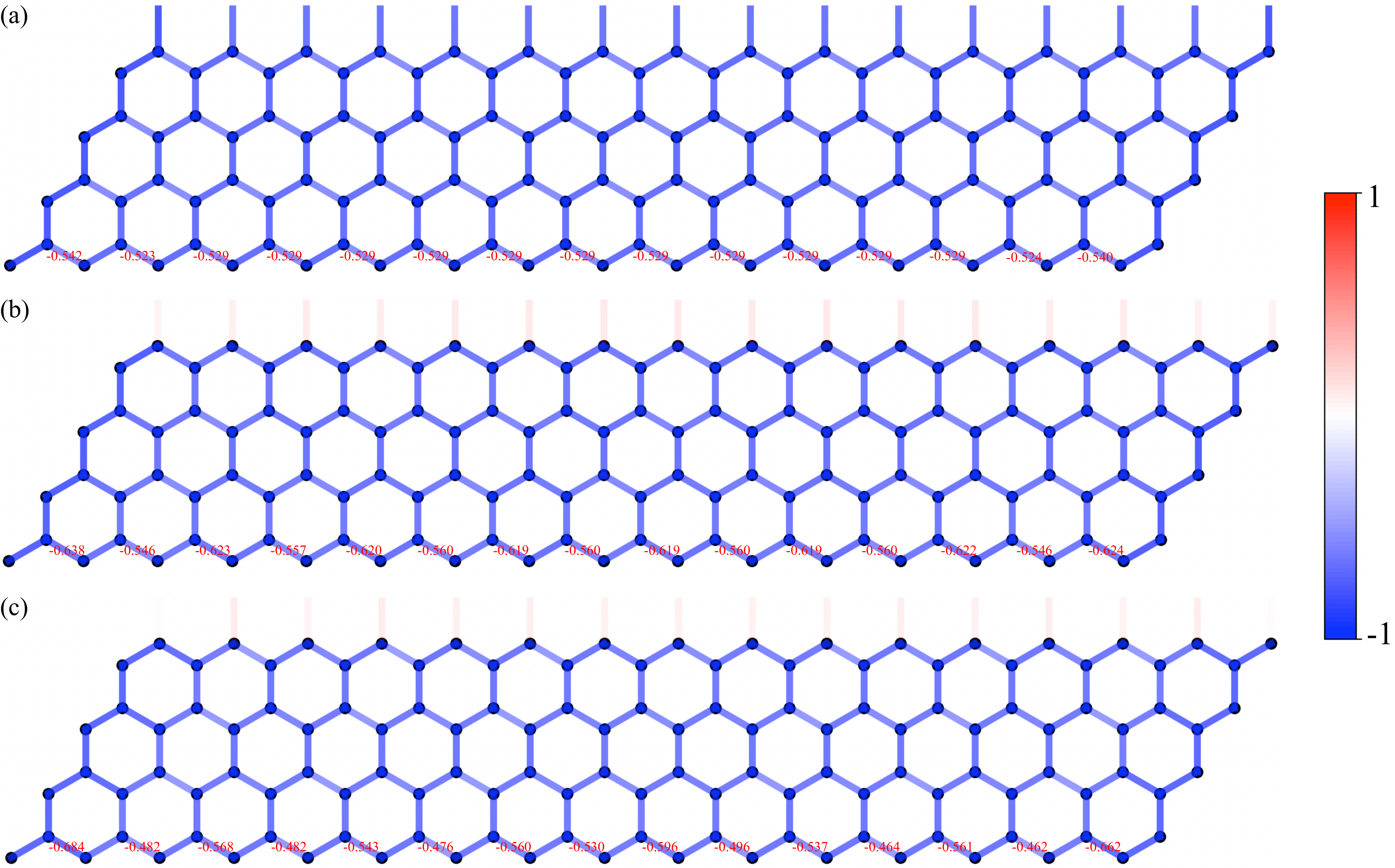}
	\caption{The expectation values of correlation function $\Lambda_{ij}$  on the NN bonds (a) on a YC-$8$ cylinder with spin flux $\theta=0$ and (b) on a twisted YC-$4$ cylinder with $\theta=\pi$. (c) $\Lambda_{ij}$ for the Gutzwiller projected $\pi$-flux state with two Dirac cones on a YC-$8$ cylinder.}\label{fig:energy_bond_yc8}
\end{figure}

Similar results can be found on YC-$8$ cylinders, see Fig.~\ref{fig:energy_bond_yc8}. The ground state obtained by DMRG with spin flux $\theta=0$ also exhibit uniform $\Lambda_{ij}$. And with $\theta=\pi$, the twisted boundary can somehow enhance the two-fold oscillation on the YC-$4$ cylinders, which can be still smeared out in the central region of cylinder. We also shown $\Lambda_{ij}$ for the Gutzwiller projected $\pi$-flux state with two Dirac cones, e.g., $\Phi_{1}=0$ and $\Phi_{2,3,4}=\pi$. Obviously, due to the boundary effect, $\Lambda_{ij}$ of a translational invariant state even has strong even-odd effects. The difference of $\Lambda_{ij}$ between $\theta=0$ and $\theta=\pi$ systems implies two different ways of cutting Dirac points. Here the MPS bond dimensions are $\chi=8000$, $12000$, and $20000$ for computing  Figs.~\ref{fig:energy_bond_yc4}(a), (b), and (c), respectively.

\section{SU(4) algebra and Spin flux}
The three generators for the Carton subalgebra of SU(4) Lie algebra are defined as
\begin{equation}
	\begin{split}
		&\lambda^1=\frac{1}{\sqrt{2}}\left(\begin{array}{cccc}
			+1 & & & \\
			& +1 & & \\
			& & -1 & \\
			& & & -1 
		\end{array}\right),~
		\lambda^2=\frac{1}{\sqrt{2}}\left(\begin{array}{cccc}
			+1 & & & \\
			& -1 & & \\
			& & +1 & \\
			& & & -1 
		\end{array}\right),~
		\lambda^3=\frac{1}{\sqrt{2}}\left(\begin{array}{cccc}
			+1 & & & \\
			& -1 & & \\
			& & -1 & \\
			& & & +1 
		\end{array}\right).
	\end{split}
\end{equation}
Above three operators defines the $U(1)\times{}U(1)\times{}U(1)$ subgroup of SU(4) group, and their eigenstates define the basis of SU(4) fundamental representation associated with the following good quantum number:
\begin{equation}
	\begin{split}
		&\left(\lambda^1,\lambda^2,\lambda^3\right)|m=1\rangle=\frac{1}{\sqrt{2}}\left(+1,+1,+1\right)|1\rangle,\\
		&\left(\lambda^1,\lambda^2,\lambda^3\right)|m=2\rangle=\frac{1}{\sqrt{2}}\left(+1,-1,-1\right)|2\rangle,\\
		&\left(\lambda^1,\lambda^2,\lambda^3\right)|m=3\rangle=\frac{1}{\sqrt{2}}\left(-1,+1,-1\right)|3\rangle,\\    
		&\left(\lambda^1,\lambda^2,\lambda^3\right)|m=4\rangle=\frac{1}{\sqrt{2}}\left(-1,-1,+1\right)|4\rangle.\\
	\end{split}
\end{equation}

And we can define the raising and lowering operators for SU(4) model as
\begin{equation}
	\begin{split}
		&\lambda^{12}=\left(\lambda^{21}\right)^\dag=\left(\begin{array}{cccc}
			0 & 1 & & \\
			& 0 & & \\
			& & 0 & \\
			& & & 0 
		\end{array}\right),\quad
	      \lambda^{13}=\left(\lambda^{31}\right)^\dag=\left(\begin{array}{cccc}
			0 &  & 1 & \\
			& 0 & & \\
			& & 0 & \\
			& & & 0 
		\end{array}\right),\quad
 	     \lambda^{14}=\left(\lambda^{41}\right)^\dag=\left(\begin{array}{cccc}
			0 & & & 1 \\
			& 0 & & \\
			& & 0 & \\
			& & & 0 
		\end{array}\right),\\
		&\lambda^{23}=\left(\lambda^{32}\right)^\dag=\left(\begin{array}{cccc}
			0 & & &  \\
			& 0 & 1 & \\
			& & 0 & \\
			& & & 0 
		\end{array}\right),\quad 
		\lambda^{24}=\left(\lambda^{42}\right)^\dag=\left(\begin{array}{cccc}
			0 & & &  \\
			& 0 & & 1\\
			& & 0 & \\
			& & & 0 
		\end{array}\right),\quad
	    \lambda^{34}=\left(\lambda^{43}\right)^\dag=\left(\begin{array}{cccc}
			0 & & &  \\
			& 0 & & \\
			& & 0 & 1\\
			& & & 0 
		\end{array}\right).
	\end{split}
\end{equation}

In order to insert a spin flux associated with the $m=1$ flavor, it is better to rewrite the lowering and raising operators at site $i$ in terms of fermionic partons:
\begin{equation}
	\lambda^{mn}_{i}=f^\dagger_{i,m}f_{i,n}.
\end{equation}
Then we can adapt a twisted boundary condition only for $f_{i,m=1}$ partons as
\begin{equation}
f_{i,1}\rightarrow{}e^{i\theta/2}f_{i+L_y{}\hat{r}_y,1}.
\end{equation}
Consequently, the physical spin operators are transformed as 
\begin{equation}
	\begin{array}{lcr}
		\lambda^{1m}_{i}\rightarrow{}e^{-i\theta/2}\lambda^{1m}_{i+L_y\hat{r}_y}, &  & {\rm\ for \ }m=2,3,4, \\
		\lambda^{m1}_{i}\rightarrow{}e^{i\theta/2}\lambda^{m1}_{i+L_y\hat{r}_y}, &  & {\rm\ for \ }m=2,3,4, \\ 
		\lambda^{mn}_{i}\rightarrow{}\lambda^{mn}_{i}, &  & {\rm\ otherwise. \ }
	\end{array}
\end{equation}
Then the Hamiltonian with a spin flux $\theta$ reads
\begin{equation}
	\begin{split}
	H(\theta)=&\sum_{\langle{}ij\rangle}\left\{\frac{1}{2}\left(\lambda^{1}_i\lambda^{1}_j+\lambda^{2}_i\lambda^{2}_j+\lambda^{3}_i\lambda^{3}_j\right)+\sum_{n=2}^{4}\sum_{m>n}\left(\lambda^{nm}_i\lambda^{mn}_j+h.c.\right) \right\}+\\
	                   &\sum_{\langle{}ij\rangle\notin{}{\rm YB}}\sum_{m=2}^{4}\left(\lambda^{m1}_i\lambda^{1m}_j+h.c.\right) +\sum_{\langle{}ij\rangle\in{}{\rm YB}}\sum_{m=2}^{4}\left(e^{i\theta}\lambda^{m1}_i\lambda^{1m}_j+h.c.\right),
	\end{split}
\end{equation}
where $\langle{}ij\rangle\in{\rm YB}$ means the NN bonds along the $y$ (periodic) boundary (for instance, $\langle{}i=1,j=8\rangle$, $\langle{}i=9,j=16\rangle$, ... shown in Fig.~\ref{fig:YC}) and $\langle{}ij\rangle\notin{\rm YB}$ the other NN bonds. By choosing $\theta=\pi$, we can obtain the Hamiltonian described by Eq.~(5) in the main text.

\end{widetext}